\documentclass{jpsj2}
\title{The Strong Coupling Fixed-Point Revisited}
\author{Author \textsc{Name}$^{1}$\thanks{Multiple authors and affiliations corr
espond using arabic numerals each other.}, Author \textsc{Name}$^{2}$\thanks{E-m
ail address: abc@def.com} and Author \textsc{Name}$^{3}$\thanks{Present address:
 Department of Applied Physics, University of Tokyo, Tokyo.}}

\inst{$^{1}$Affiliation 1 \\
$^{2}$Affiliation 2 \\
$^{3}$Affiliation 3}

\author{ \textsc{A.C. Hewson}$^{}$\thanks{Email address: a.hewson@imperial.ac.uk}}

\inst {Dept. of Mathematics, Imperial
College, London SW7 2BZ, UK. }

\abst{In recent work we have  shown that  the Fermi liquid aspects of the strong coupling fixed point of
 the
s-d and Anderson models can brought out more clearly by interpreting the fixed point as a
renormalized Anderson model, characterized by a renormalized level
 $\tilde\epsilon_d$,  resonance width, $\tilde\Delta$,
 and interaction $\tilde U$, and a simple prescription for their 
 calculation was given using  the numerical renormalization group (NRG).
These three parameters completely specify a renormalized 
 perturbation theory (RPT) which leads to exact
 expressions for the low temperature behaviour. Using a combination of the two
 techniques, NRG to determine  $\tilde\epsilon_d$, $\tilde\Delta$,
 and  $\tilde U$, and then substituting these in the RPT
 expressions gives a very efficient and accurate way of calculating the
low temperature  behaviour of the impurity as it avoids the necessity  of subtracting out the
conduction electron component.
Here we extend this approach to an Anderson model in a magnetic field,
so that  $\tilde\epsilon_d$,  $\tilde\Delta$,
 and  $\tilde U$ become dependent on the magnetic field. The
 de-renormalization  of
 the renormalized quasiparticles can then be followed as  the magnetic
field strength is increased. Using these running coupling constants in
a RPT calculation we derive  an expression for the low temperature
conductivity for arbitrary magnetic field strength.\par}

\kword{Strong coupling fixed point, Anderson model, numerical renormalization group,
renormalized perturbation theory, Fermi liquid.}

\begin{document}
\maketitle

\section{Introduction} 
After Wilson's seminal numerical renormalization group (NRG) solution of the
spin 1/2 s-d model  \cite{wil} it was very soon recognized by Nozi\`eres \cite{noz} that the low energy strong coupling fixed point corresponds to  Fermi liquid behaviour. Using a phenomenological description of
the phase shift Nozi\`eres  then gave an analytic derivation of  Wilson's result 
for the "$\chi/\gamma$" ratio $R$\cite{com}, $R=2$,  and derived the leading form of the low temperature conductivity in terms of the Kondo temperature \cite{noz}. The important characteristic of a Fermi liquid is the 1-1 correspondence of the single particle excitations with those of the non-interacting system.
This correspondence does not apply in the case of the s-d model because the model has a constraint of a fixed occupancy $n_d$ of the impurity site, $n_d=1$ for $S=1/2$, which
implies  a strong local interaction to enforce it. 
The 
Anderson model,  however, which is equivalent to the s-d model ($S=1/2$) in the local moment 
regime, is truly non-interacting
when the interaction term $U$ is set to zero, so the 1-1 correspondence of the single particle excitations with those of the non-interacting system should hold, even in the local moment limit. 
This implies that low energy fixed point of the  s-d model and Anderson model should be described more naturally
in terms of a renormalized Anderson model.   
The  Anderson model \cite{am} has the form,
\begin{equation} H_{\rm AM}=\sum\sb {\sigma}\epsilon\sb {d}
d\sp {\dagger}\sb {\sigma}
d\sp {}\sb {\sigma}+
Un\sb {d,\uparrow}n\sb {d,\downarrow}
 +\sum\sb {{ k},\sigma}( V\sb { k}d\sp {\dagger}\sb {\sigma}
c\sp {}\sb {{ k},\sigma}+ V\sb { k}\sp *c\sp {\dagger}\sb {{
k},\sigma}d\sp {}\sb {\sigma})+\sum\sb {{
k},\sigma}\epsilon\sb {{ k},\sigma}c\sp {\dagger}\sb {{ k},\sigma}
c\sp {}\sb {{
k},\sigma},\label{ham}\end{equation}
where $\epsilon_d$ is the energy of the impurity level, $U$ the interaction at the impurity site,
 and $V_{k}$ the hybridization matrix element to a band of conduction electrons with
energy $\epsilon_k$.When $U=0$ the local level broadens
into a resonance, corresponding to a localized quasi-bound state,
whose width depends on the quantity $ \Delta(\omega)=\pi\sum\sb {k}|
V\sb {k}|\sp 2\delta(\omega -\epsilon\sb { k})$. It is usual to
consider the case of a wide conduction band with a flat density of
states where $\Delta(\omega)$ becomes independent of $\omega$ and can
be taken as a constant $\Delta$.\par 
In the wide conduction band limit the one-electron Green's function, $G_{d}(\omega)$
takes the form
\begin{equation}G_{d}(\omega)={1\over \omega-\epsilon_d+i\Delta-\Sigma(\omega)}\label{gf}\end{equation}
where $\Sigma(\omega)$ is the self-energy and $\Delta$ is the width of the resonance
at $\epsilon_d$ when $U=0$.
 Near the Fermi level for small $\omega$,
$\Sigma(\omega)=\Sigma(0)+\omega\Sigma'(0)+{\rm O}(\omega^2)$, and if this is substituted into
(\ref{gf}) then for small $\omega$ the denominator is of the same form as that for the non-interacting
system with a renormalized level $\tilde\epsilon_d$ and resonance width $\tilde\Delta$ given by
 \begin{equation}\tilde\epsilon_{\rm d}=z(\epsilon_{\rm d}
+\Sigma_{}(0)),\quad\tilde\Delta =z\Delta ,\label{ren1}\end{equation}
where $z$,  the wavefunction renormalization
 factor, is given by
$z={1/{(1-\Sigma'_{}(0))}}$, and the prime indicates a derivative with respect to $\omega$. 
The 1-1 correspondence is evident when one calculates the quasiparticle occupation number 
$\tilde n_d$ at $T=0$,
\begin{equation} \tilde n_{d\sigma}={1\over
2}-{1\over\pi}\tan ^{-1}\left
({\tilde\epsilon_{{\rm d}}}\over{\tilde\Delta}\right
)=n_{d\sigma},\label{qpfsr}\end{equation} 
which is equal to the impurity occupation number $n_{d\sigma}$ for spin $\sigma$ at $T=0$ from the Friedel sum rule
\cite{fsr}. The corresponding quasiparticle density of states $\tilde\rho_{d}(\omega)$
is given by
\begin{equation}\tilde\rho_{d}(\omega)={\tilde\Delta/\pi\over (\omega-\tilde\epsilon_d)^2+\tilde\Delta^2}.\label{qpdos}\end{equation}
It follows from Fermi liquid theory that the impurity specific heat coefficient $\gamma$, as calculated from
these non-interacting quasiparticles, is exact as $T\to 0$, and is given by
\begin{equation}\gamma_{\rm imp}={2\pi^2\over 3}\tilde\rho_{d}(0).\label{rgam}\end{equation}
In the original NRG calculations for the Anderson model the low energy fixed point was
analysed as a strong coupling $V\to \infty$ fixed point \cite{KWW}. In this limit the impurity is decoupled,
and so the analysis does not bring out the 1-1  correspondence of the single particle excitations with those of the original Anderson model. Recently we (Hewson, Oguri and Meyer\cite{hom}) have reanalysed the strong coupling fixed point as
a $U=0$ fixed point with a finite $V$,  which leads directly to the renormalized
parameters $\tilde\epsilon_d$ and
$\tilde\Delta$. 
The NRG calculations are based on linear chain  form for the  Anderson model, with a 
discretized conduction electron spectrum and a
discretization parameter $\Lambda>1$. Starting with the impurity at the end of the chain,
the Hamiltonians corresponding to a finite lengths of chain are diagonalized
 iteratively,   adding a new  site to the chain with each iteration.
If the lowest energy single-particle $E_p(N)$ or single-hole excitations $E_h(N)$ of the interacting Anderson model
for a chain with $N+2$ sites can be described by a non-interacting renormalized Anderson model
 then $E=E_p(N)$  or $E=E_h(N)$   should satisfy the equation,
\begin{equation}{E\Lambda^{-(N-1)/2}}-\tilde\epsilon_d(N)=\Lambda^{(N-1)/2}\tilde V(N)^2g_{00}(E),\label{poles}\end{equation}
where the effective parameters, $\tilde\epsilon_d(N)$ and $\tilde\Delta(N)=\pi\tilde V(N)^2/D$,
should be  independent of $N$. The function $g_{00}(\omega)$ is the local Green's function  for the first  conduction electron site of the chain when decoupled from the
impurity ($V=0$) (see \cite{hom} for details).

\begin{figure}[tb]
\begin{center}
 \includegraphics[width=0.6\textwidth]{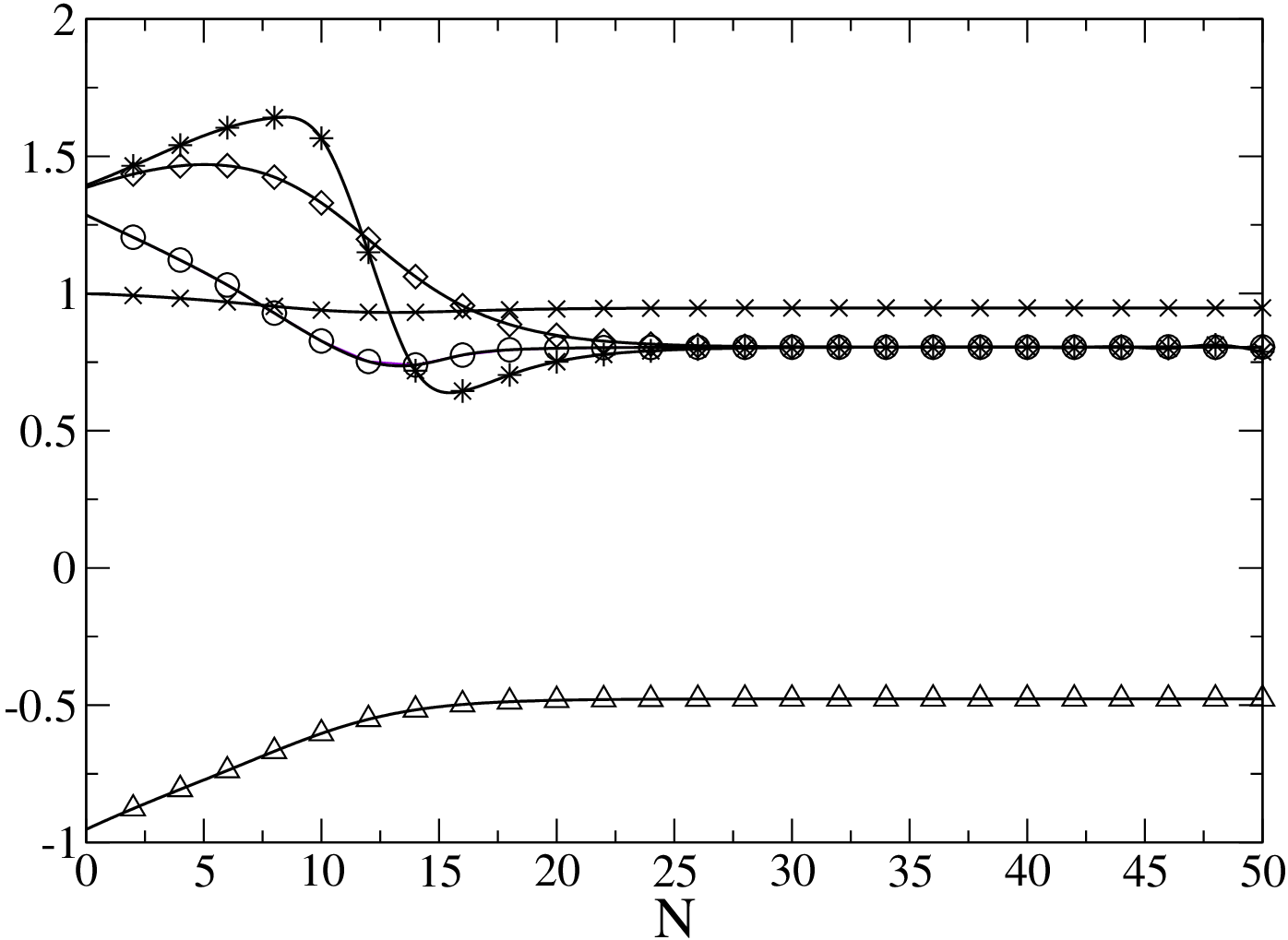}
\end{center}
\caption{Plots  of the parameters,  $\tilde\Delta(N)/\Delta$ (crosses), $\tilde\epsilon_d(N)/\pi\Delta$ (triangles),
 $\tilde U_{pp}(N))/\pi\Delta$ (diamonds), $\tilde U_{hh}(N)/\pi\Delta$ (stars)
 and $\tilde U_{ph}(N)/\pi\Delta$ (circles), with $N$,
for the Anderson model with bare parameters, $U=0.04$, $\pi\Delta=0.03$,  and
 $\epsilon_d=-0.05$. In this and subsequent figures the full lines are
 interpolations using the calculated data points.}
\label{figure1}
\end{figure}

In Figs. 1 and 2
we present results for  $\tilde\epsilon_d(N)$ and $\tilde\Delta(N)$, determined by these two equations  as a function of $N$. In the first case shown in figure 1 we take a relatively
weak coupling example with bare parameters, $\pi\Delta=0.03$, $\epsilon_d=-0.05$ and $U=0.04$ (bandwidth $2D=2$ in all cases), which is such that $U\rho_{d,{\rm mf}}(0)= 0.331< 1$, where $\tilde\rho_{d,{\rm mf}}(0)$ is the mean field density of states at the Fermi level, $\tilde\rho_{d,{\rm mf}}(0)=\Delta/\pi(\tilde\epsilon_{d,{\rm mf}}^2+\Delta^2)$ with $\tilde\epsilon_{d,{\rm mf}}=\epsilon_d+Un_{d,{\rm mf}}/2$,  and so does not satisfy the mean field (Hartree-Fock) criterion for a local moment. The renormalized parameters are independent of $N$ for $N>26$, which confirms that these single-particle excitations
on the lowest energy scale can indeed be described by a renormalized Anderson model.
The asymptotic value of $\tilde\Delta/\Delta=0.947 $ for large $N$, and so
$\tilde\Delta$ differs little from its bare value. The value of
$\tilde\epsilon_d=-0.0143$, and is very approximately of the same order as that predicted
from mean field theory $\tilde\epsilon_{d,{\rm mf}}=-0.0166$.\par

\begin{figure}[tb]
\begin{center}
\includegraphics[width=0.6\textwidth]{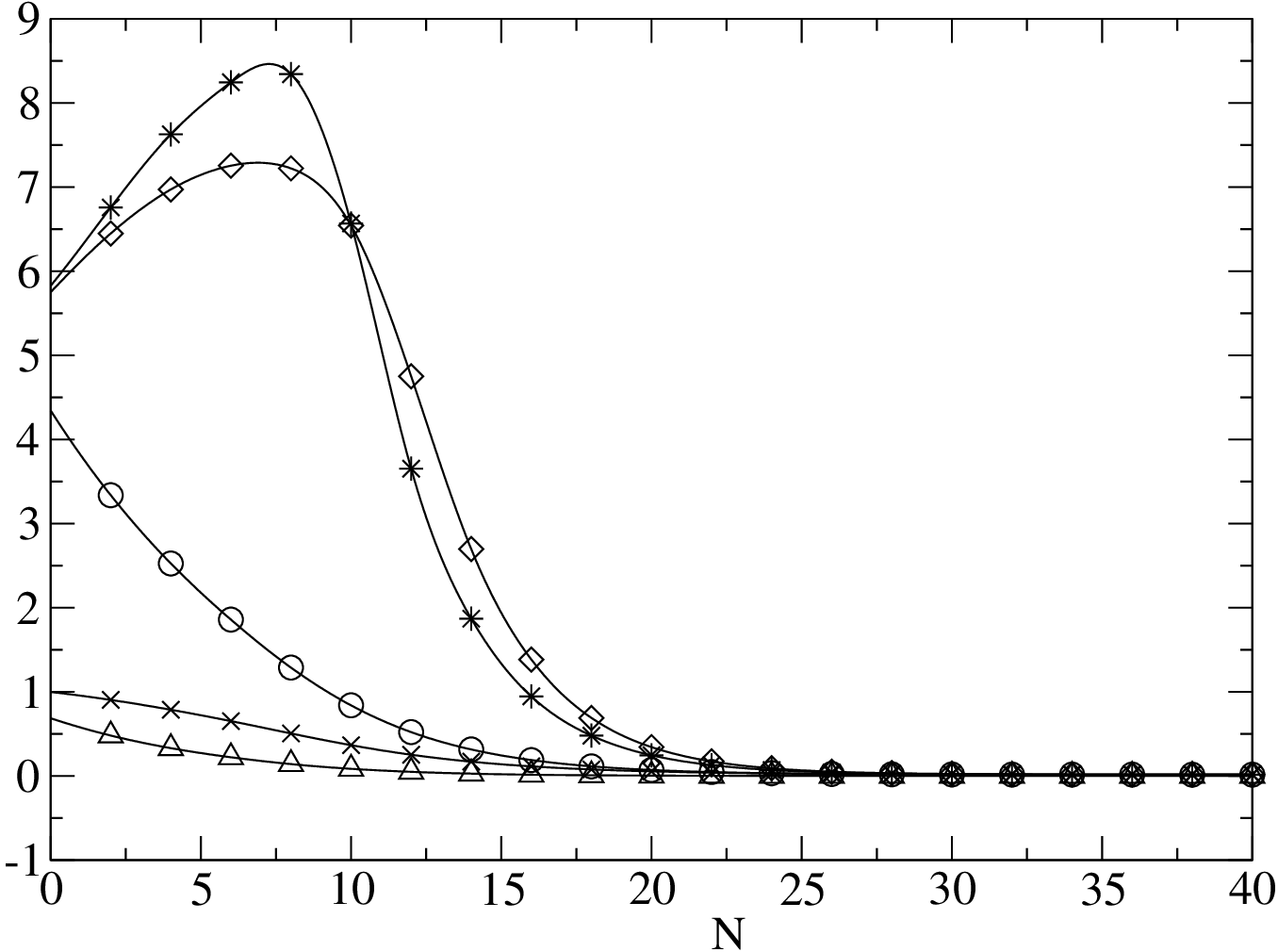}
\end{center}
\caption{Plots  of the parameters,  $\tilde\Delta(N)/\Delta$ (crosses), $\tilde\epsilon_d(N)/\pi\Delta$ (triangles),
 $\tilde U_{pp}(N))/\pi\Delta$ (diamonds), $\tilde U_{hh}(N)/\pi\Delta$ (stars)
 and $\tilde U_{ph}(N)/\pi\Delta$ (circles), with the $N$,
for the Anderson model with bare parameters, $U/\pi\Delta=5.0$, $\pi\Delta=0.03$,  and
 $\epsilon_d=-0.05$
.}
\label{figure2}
\end{figure}

In Fig. 2 we give the corresponding results for $U=0.15$,
keeping the other parameters the same. For this value of $U$ mean field theory
would predict the breaking of local spin symmetry as $U\tilde\rho_{d,{\rm mf}}(0)=
2.02>1$,
and hence this case corresponds to a system with a local moment and can be
described by an effective s-d model. We see a considerable renormalization such
that $\tilde\Delta/\Delta=0.0158$, and $\tilde\epsilon_d$ becomes very small
$\tilde\epsilon_d=1.06\times 10^{-5}$ so the effective level is very close to the
Fermi level, as to be expected in the almost localized limit $n_{d}\approx 1$.
Having determined $\tilde\epsilon_d$ and $\tilde\Delta$, the impurity occupation and
the specific heat coefficient $\gamma$ can be determined from eqs. (\ref{qpfsr}) and (\ref{rgam}).

 The renormalized quasiparticles must interact with one another, and this interaction must come
into play as soon as two or more single particle excitations are created from
the interacting ground state. 
  If the lowest two-particle excitation from the ground state for the interacting
system
for a given $N$ has an energy $E_{pp}(N)$, then we can calculate $\tilde U$ by equating the
energy difference $E_{pp}(N)-2E_{p}(N)$ to that calculated by adding  an local
interaction  term to the  effective Anderson model for the non-interacting quasiparticles\cite{KWW,hom}.
  For finite $N$ we can use this equation to define an
$N$-dependent
renormalized interaction $\tilde U_{pp}(N)$,
   \begin{equation}E_{pp}(N)-2E_{p}(N)=\tilde U(N)\Lambda^{(N-1)/2}
|\psi_{p,1}^*(-1)|^2|\psi_{p,1}^*(-1)|^2,
\label{tilu}\end{equation}
where $|\psi_{p,1}(-1)|^2$ is given by
\begin{equation}
 |\psi_{p,1}|^2= {1\over 1-\tilde V^2(N)\Lambda^{(N-1)}{g'}_{00}(E_p(N))},
\label{UN}
\end{equation}
where ${g'}_{00}(\omega)$ is the derivative of $g_{00}(\omega)$.\par
Alternatively we could consider the same procedure for a two hole excitation $E_
{hh}(N)$
and in a similar way define an $N$-dependent
renormalized interaction $\tilde U_{hh}(N)$, or a particle-hole excitation
 $E_{ph}(N)$ to define a
renormalized interaction $\tilde U_{ph}(N)$. In this latter case, as a positive
$U$ leads to  particle-hole attraction, we use $E_p(N)+E_h(N)-E_{ph}(N)$ on the
left-hand
side of eq. (\ref{tilu}).\par
If these two particle excitations can be described by an effective Anderson model then
$\tilde U_{pp}(N)$, $\tilde U_{hh}(N)$ and $\tilde U_{ph}(N)$ should be independent of
$N$ and also independent of the particle and hole labels.  The values of $\tilde U_{pp}(N)$,
$\tilde U_{hh}(N)$ and $\tilde U_{ph}(N)$, for the case $U=0.04$ as a function
of $N$ are shown in Fig. 1, and those for $U=0.15$ in Fig. 2, where
$\pi\Delta=0.03$ and $\epsilon_d=-0.04$ in both cases. The three interaction
parameters
can be seen to converge to a unique value for large $N$  both in the weak
and strong coupling cases.  \par

In Fig. 3 the renormalized interaction parameters in the strong coupling case  for $N>25$ of the plot 
given in Fig. 2 are
shown over a smaller  energy range. 
It  can  be seen  that  $\tilde U_{pp}(N)$,
$\tilde U_{hh}(N)$ and $\tilde U_{ph}(N)$ not only have a unique limit but
that this  limit coincides with the asymptotic value of $\pi\tilde\Delta(N)$
for large $N$, so that $\pi\tilde\Delta=\tilde U$; consequently there is only one
effective parameter scale in the localized (Kondo) regime.\par
\par

\begin{figure}[tb]
\begin{center}
\includegraphics[width=0.6\textwidth]{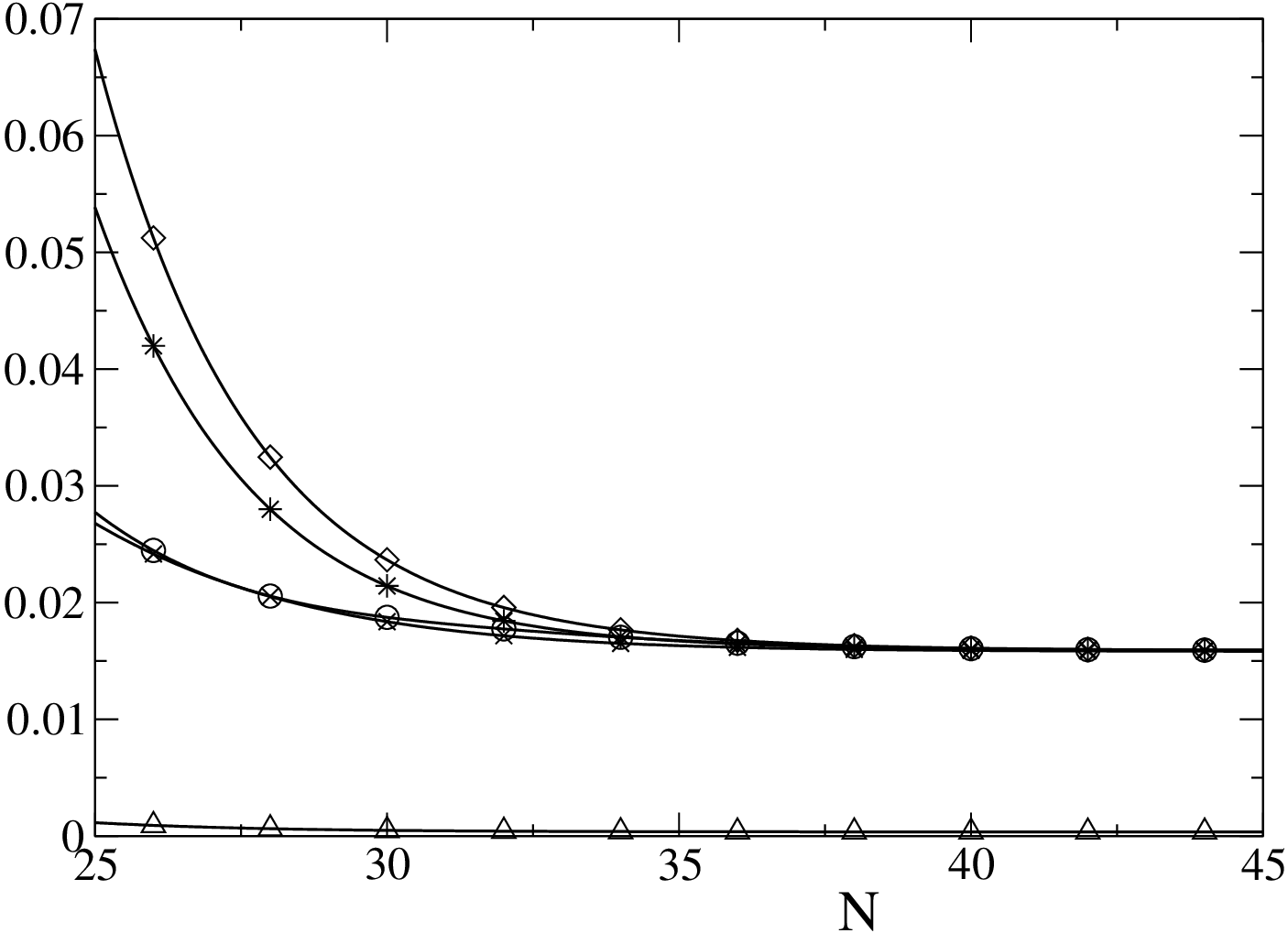}
\end{center}
\caption{The same plot as in Fig. 2  for  $N>25$  but shown over a smaller energy range.
   The parameters correspond to the almost localized (Kondo) limit; the renormalized level $\tilde\epsilon_d/\pi\Delta$ (triangles)
is very close to the Fermi level, and  the limiting value of
  $\tilde\Delta(N)/\Delta$ for large $N$
  (crosses) coincides  with the limiting
  values of $\tilde U_{pp}(N))/\pi\Delta$  (diamonds), $\tilde U_{hh}(N)/\pi\Delta$ 
 and $\tilde U_{ph}(N)/\pi\Delta$ (circles).  }
\label{figure3}
\end{figure}
\section{Renormalized Perturbation Theory}
An alternative renormalization approach to the Wilson technique is the renormalized perturbation theory (RPT)\cite{ft} as originally developed to deal with the
infinities arising in quantum electrodynamics (QED). While the Wilson approach is 
based on the progressive elimination the higher order
excitations  to obtain an effective low energy model, the renormalized perturbation theory
is essentially a reorganization of perturbation theory  so that a perturbation
expansion can be carried for the same model, but in terms of 
 parameters appropriately renormalized for very low energy scales.  These low energy scales are where almost all observations are made in QED so that
the renormalized parameters can be taken from experiment. As the RPT is not simply about
the cancellation  of infinities, but working with parameters appropriate to the energy scale
under investigation, it can be applied quite generally. This approach has been
developed for the Anderson model, and leads naturally to the quasiparticle
description\cite{rpt1,rpt2}. Here we briefly review some of the main results which will be required
in the later sections of this paper. The renormalized parameters
of the non-interacting quasiparticles are a renormalized level
$\tilde\epsilon_d$ and resonance width $\tilde\Delta$,  as given in eq. (\ref{ren1})
in terms of the self-energy and its derivative at $T=\omega=0$. The renormalized local interaction
$\tilde U$ is identified with the fully dressed irreducible 4-vertex, $\Gamma_{\uparrow\downarrow}
(\omega_1,\omega_2,\omega_3,\omega_4)$ at $\omega_1=\omega_2=\omega_3=\omega_4=0$,
\begin{equation}\tilde U=z^2\Gamma_{\uparrow\downarrow}
(0,0,0,0),\label{ren2}\end{equation}
which is rescaled by $z^2$ so the propagator for the quasiparticles is normalized.\par
In applying the renormalized perturbation expansion the Lagrangian for the  Anderson model ${\cal L}_{\rm AM}(\epsilon_d,
\Delta,U)$ is rewritten in the form,
 \begin{equation}{\cal L}_{\rm AM}(\epsilon_d,
\Delta,U)={\cal L}_{\rm AM}(\tilde\epsilon_d,
\tilde\Delta,\tilde U) +{\cal L}_{\rm c}(\lambda_1,\lambda_2,\lambda_3).\label{rlag}\end{equation}
The first term on the right hand side is simply the Lagrangian for the Anderson model
expressed in terms of the renormalized parameters, and the second part is the
remainder or  counter term. The three parameters $\lambda_1$, $\lambda_2$ and $\lambda_3$
are fully determined order by order in the expansion in powers of $\tilde U$ by the
requirement that they prevent overcounting and cancel off any terms that further
renormalize the particles  as these quantities have be taken
to be fully renormalized already. Hence the three renormalized parameters $\tilde\epsilon_d$,
$\tilde\Delta$ and $\tilde U$ are sufficient to specify the RPT precisely.\par

The first order perturbation theory in $\tilde U$ gives results for the impurity spin  and charge
susceptibilities at $T=0$,
 \begin{equation}\chi_{s}={(g\mu_{\rm B})^2\over 2}\tilde\rho_d(0)(1+
 \tilde U\tilde\rho_d(0)),\quad\chi_{c}=2\tilde
 \rho_{d}(0)(1-\tilde
 U\tilde\rho_d(0)),\label{rsus}\end{equation}
where $\tilde\rho_d(\omega)$ is given by eq. (\ref{qpdos}). These results
  can shown to be exact by the use of a Ward identity, and are equivalent to the results as
first derived by Yamada\cite{yam} from an analysis of a perturbation expansion in powers
 of $U$.
The `$\chi/\gamma$' ratio $R$ is then given by $R=1+\tilde U\tilde\rho_d(0)$.\par
The  RPT result for the
renormalized self-energy  to second order in $\tilde U$ 
for the symmetric model gives the exact low temperature result for the
 conductivity  $ \sigma(T)$ to second order in the temperature $T$, which is
 given by 
\begin{equation} \sigma(T)=\sigma_0\left\{1+{\pi^2\over
    3}\left(T\over\tilde\Delta\right)^2\left[1+2\left({\tilde
    U\over\pi\tilde\Delta}\right)^2\right]+{\rm
    O}(T^4)\right\}.\label{sigT}\end{equation}
     When the renormalized parameters
    are expressed in terms of the self-energy and the vertex function
    these results coincide with the exact expressions derived by
    Yamada and Yosida \cite{yam} from an analysis of perturbation theory to all
    orders $U$, and in the localized regime ($\tilde U/\pi\tilde\Delta\to 1$) with the Fermi-liquid results of Nozi\`eres \cite{noz}. The corresponding result for the differential conductance through
a quantum dot to second order in the applied voltage has been given by Oguri\cite{oguri}.\par

Higher order  terms in $\tilde U$ can be used to estimate the leading corrections to the 
Fermi liquid results\cite{rpt2}. 
Estimates of the induced magnetization $M(h)$
as a function of magnetic field $H$, where $h=g\mu_{\rm B}H/2$ to order $H^3$ have been made using RPT to third order
in $\tilde U$ for the particle-hole symmetric model. The result to this order is
\begin{equation}
M(h)={g\mu_{\rm B}\over\pi}\left\{\left(1+{\tilde U\over \pi\tilde\Delta}\right)\left({ h\over\tilde\Delta}\right)
+\left[1+4\left({\tilde U\over \pi\tilde\Delta}\right)+A\left({\tilde U\over \pi\tilde\Delta}\right)^2
+B\left({\tilde U\over \pi\tilde\Delta}\right)^3\right]\left({ h\over\tilde\Delta}\right)^3+..\right\}\end{equation}
where  the coefficients $A$ and $B$ are given by
\begin{equation} A={38\over 3}-{3\pi^2\over 4},\quad{\rm and}\quad B={644\over 9}-7\pi^2.\end{equation}

Though this estimate of the coefficient of the $H^3$ term is not exact, it  is asymptotically exact for small $U$ ($U\to 0$) and  
differs at the most by 4\% from the Bethe ansatz result in the localized Kondo regime
($U\gg\pi\Delta$).\par
In principle it should be possible to derive results appropriate for all energy scales
for the RPT given the three renormalized parameters, as nothing is neglected  when the counter terms are taken 
into account. However, higher order calculations get progressively more difficult and it is
unlikely that the summation of any subset of terms is likely to provide reasonably accurate
results on all energy scales.  As we shall discuss later, however, it might prove possible to
calculate a set of running coupling constants, appropriate to the energy scale under consideration, and use the low order RPT to cover all energy scales in this way.\par
\section{Renormalized Anderson Model}
  The first term on the right hand side of eq. (\ref{rlag}) can be identified as the fixed point of the Wilson NRG approach,
because as $T\to 0$ the effect of the counter terms is to normal order the interaction term
so that it only comes into play when two or more excitations are created from the ground state.
The Hamiltonian for the renormalized Anderson model, which describes the behaviour near the low
energy fixed point, therefore, can be written as
\begin{equation} \tilde H_{\rm AM}=\sum\sb {\sigma}\tilde\epsilon\sb {d}
d\sp {\dagger}\sb {\sigma}
d\sp {}\sb {\sigma}+
\tilde U : n\sb {d,\uparrow}n\sb {d,\downarrow}:
 +\sum\sb {{ k},\sigma}(\tilde V\sb { k}d\sp {\dagger}\sb {\sigma}
c\sp {}\sb {{ k},\sigma}+\tilde V\sb { k}\sp *c\sp {\dagger}\sb {{
k},\sigma}d\sp {}\sb {\sigma})+\sum\sb {{
k},\sigma}\epsilon\sb {{ k},\sigma}c\sp {\dagger}\sb {{ k},\sigma}
c\sp {}\sb {{
k},\sigma},\label{rham}\end{equation}
where  the colon brackets indicate that the expression within them must be normal-ordered. 
This renormalized model is similar to that used in earlier phenomenological local Fermi-liquid theories \cite{nh}, but here it also includes a quasiparticle interaction term.\par

We can identify  $\tilde U$ as defined from eq. (\ref{ren2}) with that from the
NRG calculation in eq. (\ref{tilu}). The NRG results for $\tilde\epsilon_d$, $\tilde\Delta$ and $\tilde U$ can then be substituted in eqs. (\ref{rsus}) to evaluate the $T=0$ spin and charge
susceptibilities. This is considerably simpler than the method used originally to evaluate these
and is very accurate as it does not involve subtracting out the conduction electron
component. It also follows analytically from these results that in the Kondo regime,
$\chi_c(0)\to 0$ and $n_d\to 1$ that $\epsilon_d\to 0$ and $\tilde U=\pi\tilde \Delta$,
as noted earlier in the results shown in Fig. 3. It is equivalent to  Nozi\`eres'
argument\cite{noz} and gives the Wilson ratio $R=2$. If we define the Kondo temperature $T_{\rm K}$
via $T_{\rm K}=(g\mu_{\rm B})^2/4\chi_s(0)$
then $\tilde U=\pi\tilde \Delta=4T_{\rm K}$.\par 
\begin{figure}[tb]
\begin{center}
\includegraphics[width=0.6\textwidth]{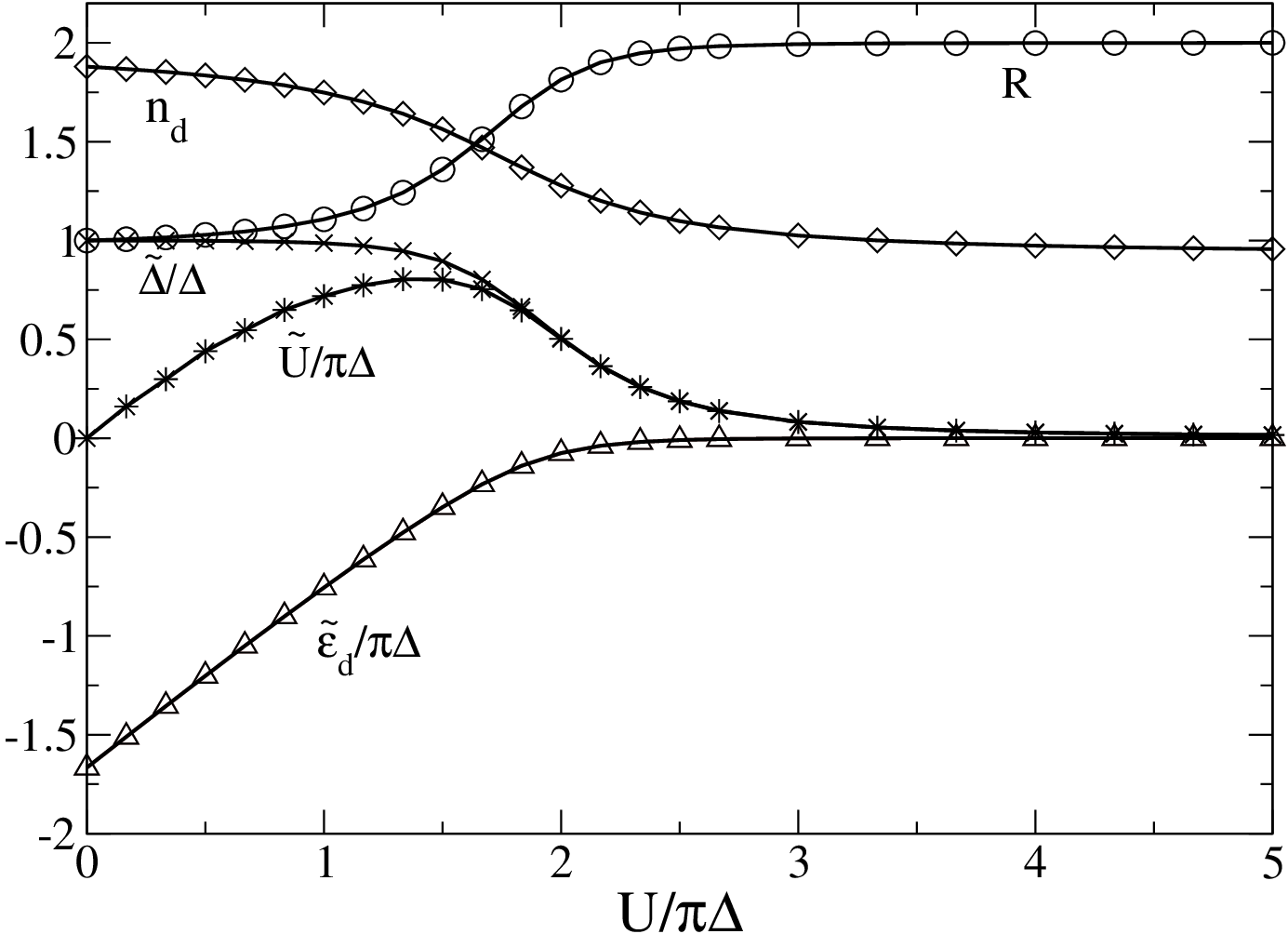}
\end{center}
\caption{A plot of the renormalized parameters $\tilde\epsilon_d$,
  $\tilde\Delta$ and $\tilde U$ as a function of the bare interaction $U$ for
  an Anderson model with $\pi\Delta=0.03$ and $\epsilon_d=-0.04$. Over this
  range of $U$ the system moves from a full orbital regime $\epsilon_d+U\ll 0$, through an
  intermediate
valence regime $\epsilon_d+U\approx 0$, to a localized (Kondo) regime
  $\epsilon_d+U\gg 0$. Also shown are the impurity occupation number $n_d$ and
  the `$\chi/\gamma$' or Wilson ratio R.}
\label{figure4}
\end{figure}

 The advantage of the renormalized Anderson approach to describe the low energy behaviour
of non-degenerate magnetic impurities is that all the parameter regimes; weak coupling, mixed valence, strong correlation, empty and full orbital regimes, can be described precisely
within a single framework with, at the most, three renormalized parameters.
In Fig. 4 we take values of $U$ over the range $0<U\le 5\pi\Delta$, with the same
values of $\epsilon_d$ and $\Delta$ as used earlier. Over this range we move from the full
orbital regime for small $U$, through an intermediate valence regime for $\epsilon_d+U\sim
0$ and the Kondo regime for $U\gg |\epsilon_d|$.
We see that
$\tilde\epsilon_d$ increases at first approximately linearly with $U$ until  $U/\pi\Delta\sim 2.0$,
  remaining very close to the Fermi level in the Kondo regime at higher values of $U$. The renormalized
resonance width $\tilde\Delta$ decreases over the same range monotonically approaching zero in the
limit $U\to\infty$.
The quasiparticle interaction
$\tilde U$ increases at first linearly with $U$, reaching a maximum for
$U/\pi\Delta\sim 1.5$, and then decreases so that its energy scale merges with
that for $\tilde\Delta$ in the Kondo regime. The initial increase of $\tilde\epsilon_d$  with $U$ in the full orbital
regime is understandable in terms of the mean field or Hartree-Fock theory,
which is approximately valid in this regime. In the mean field theory
there is no $\omega$-dependence in the self-energy so $z=1$ and $\tilde\Delta_{\rm mf}=\Delta$.
There is an effective level given by
$\tilde\epsilon_{d,{\rm mf}}=\epsilon_d+U n_{d,{\rm mf}}/2$, where $
n_{d,{\rm mf}}$ is the total impurity occupation calculated within the
mean field theory. The value of $\tilde U$ in proportional to $U$ for very small $U$,
and for larger values of $U$, its value can be estimated from (\ref{tilu}) using perturbation theory.\cite{rpt2}   These expressions give the general trend as a
function of $U$ in the full orbital regime $U\tilde\rho_{d,{\rm mf}}\ll 1 $ but are not valid
as the strong correlation regime is approached, where $\tilde\Delta$ begins to
differ from the bare value $\Delta$, and breaks down completely
 in the Kondo regime where $\tilde\Delta$ is strongly renormalized (see reference\cite{hom}
for more extensive results).\par
Also plotted in Fig. 4 is the  Wilson ratio,
which is given by  $R=1+\tilde U\tilde\rho_d(0)$, and  the total impurity occupation number
$n_d$.   
The occupation number decreases from 1.88 for $U=0$ to almost 1 in the large
$U$ regime,
corresponding to localization of the d-electron.  The Wilson ratio  increases  from 1 in the small $U$
regime  and asymptotically approaches 2 in the Kondo regime.

 \section{De-renormalization as a Function of Magnetic Field}  
If we introduce a magnetic field $H$, we can again generalize the definition of the
renormalized parameters, $\tilde\epsilon_{d\sigma}$ 
$\tilde\Delta$ and  $\tilde U$, such that they become functions of the magnetic field $H$.
For simplicity we confine the discussion to the particle-hole symmetric model with
$\epsilon_d=-U/2$. If we absorb the zero field Hartree-Fock contribution to the self-energy, 
$\Sigma(0,0)=U/2$ then  $\epsilon_d=0$, and
$\Sigma_{\uparrow}(0,h)=-\Sigma_{\downarrow}(0,h)=-\Sigma_{\uparrow}(0,-h)$,
where $h=g\mu_{\rm B}H/2$.  
Eq. (\ref{ren1}) gets replaced by
 \begin{equation}\tilde\epsilon_{\rm d,{\sigma}}(h)=z(h)(-h\sigma
+\Sigma_{\sigma}(0,h)),\quad\tilde\Delta(h) =z(h)\Delta ,\label{renh1}\end{equation}
where $z(h)$ is given by
$z(h)={1/{(1-\Sigma'_{\uparrow}(0,h))}}$. 
The occupation of the impurity level is still given by the Friedel sum rule,
\begin{equation} n_{d\sigma}(h)={1\over 2}-{1\over\pi}{\rm tan}^{-1}\left(\tilde\epsilon_{d\sigma}(h)\over\tilde\Delta(h)\right).\label{numh}\end{equation}
The definition (\ref{tilu}) of $\tilde U$ can be straightforwardly generalized to define an
field dependent local interaction $\tilde U(h)$. 
As  $\tilde\epsilon_{d\uparrow}(h)=-\tilde\epsilon_{d\downarrow}(h)$, it will be
convenient to define a single
effective level $\epsilon_{d}(h)$ via $\tilde\epsilon_{d}(h)=-\sigma\tilde\epsilon_{d\sigma}(h)$.
The impurity magnetization  $M(h)$
at $T=0$ 
is then  given simply by 
\begin{equation} M(h)={g\mu_{\rm B}\over\pi}{\rm tan}^{-1}\left({\tilde\epsilon_{d}(h)\over\tilde\Delta(h)}\right).\label{mag}\end{equation}
The expressions given earlier for the quasiparticle density of states (\ref{qpdos}),
impurity spin and charge susceptibilities (\ref{rsus}), and specific heat coefficient (\ref{rgam})
can be generalized to include the magnetic field dependence by replacing $\tilde\epsilon_d$,
$\tilde\Delta$ and $\tilde U$, by $\tilde\epsilon_d(h)$,
$\tilde\Delta(h)$ and $\tilde U(h)$.\par
The simplest way  to calculate the field dependent renormalized parameters
using the NRG for the 
particle-hole symmetric model is to exploit the spin-isospin symmetry of the model,
which is such that the symmetric model in a magnetic field with positive-$U$ is equivalent
to a negative-$U$ asymmetric model in the absence of a magnetic field with 
$\epsilon_d+U/2=-h$ and  
spin and charge (isospin) interchanged.  Calculations can then be carried out
 in the absence of a magnetic field using the asymmetric model with negative-$U$. 
The isospin
symmetry of the particle-hole symmetric model is then effectively exploited
via the spin symmetry. This has two advantages, it 
 enables one to retain more states in the NRG iterations, and requires no modification of the
standard NRG program. \par
\begin{figure}[tb]
\begin{center}
\includegraphics[width=0.6\textwidth]{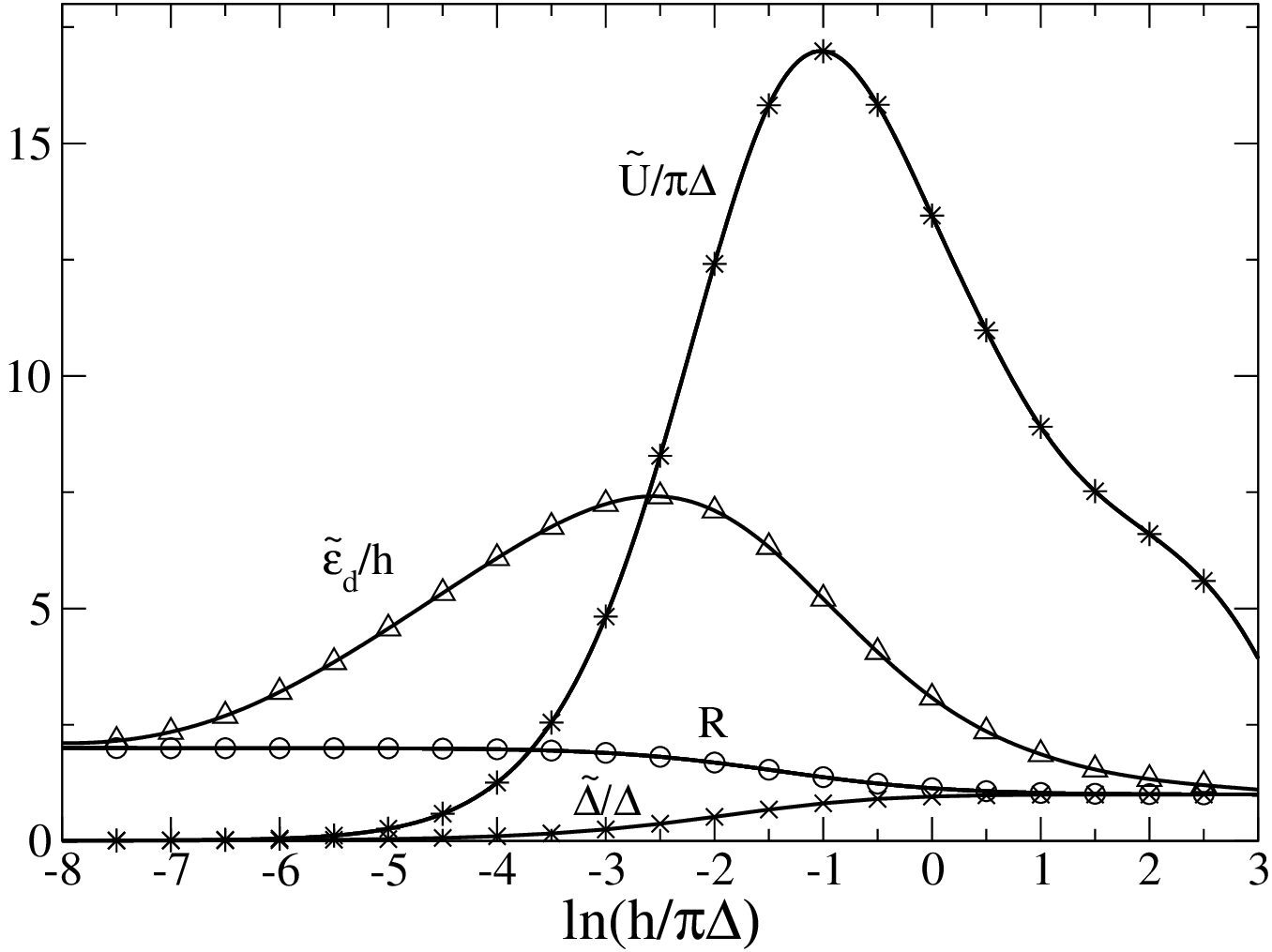}
\end{center}
\caption{The renormalized parameters
  $\tilde\epsilon_d(h)/h$. $\tilde\Delta(h)/\Delta$, $\tilde U(h)/\pi\Delta$
  and  the Wilson ratio $R(h)$ plotted as a function of the logarithm of the
  magnetic field $h/\pi\Delta$ for the symmetric Anderson model with
  $U/\pi\Delta=5$, $\pi\Delta=0.03$,  and a bandwidth $2D=2$. The parameters vary from the strong
  correlation values in weak field, $\pi\tilde\Delta(0)=\tilde U(0)$,
  $R(0)=2=\lim_{h\to 0}\tilde\epsilon_d(h)/h$, to those of the bare model, $\tilde
  U(h)=U$, $\tilde\Delta(h)=\Delta$, $R(h)=1=\tilde\epsilon_d(h)/h$ in fields
  $h\gg U$. }
\label{figure5}
\end{figure}
In Fig. 5 we give the results for $\tilde\epsilon_d(h)/h$,
$\tilde\Delta(h)/\Delta$ and $\tilde U(h)/\pi\Delta$ as a function of $h$ for the symmetric model 
with $U/\pi\Delta=5$, which corresponds to a Kondo temperature $T_{\rm K}/\pi\Delta=0.00204$.  We can follow the gradual de-renormalization of the quasiparticles as the
magnetic field increases. In weak field $\tilde\epsilon_d/h\to 2$ as $h\to 0$, so the level
splitting in a magnetic field of the local spin-up and spin-down states is twice
the Zeeman splitting for free quasiparticles. This enhancement factor of 2 is the same 
as that for the Wilson or $\chi/\gamma$ ratio. As the magnetic field strength increases
$\tilde\epsilon_d(h)/h$   increases to a maximum, and only for magnetic field strengths larger than $h\gg U$
does this ratio finally approach the free particle value,  $\tilde\epsilon_d(h)/h\to 1$. 
There is an even more dramatic increase in the value of $\tilde U(h)/\pi\Delta$.
For $h\to 0$  we are in the strong correlation or Kondo limit,  $\tilde U(0)/\pi\Delta(0)=4T_{\rm K}/\pi\Delta$, so the value of $\tilde U(h)$ is small in the weak field regime. However, it increases significantly
with an increase of field strength  to a maximum, approximately three times greater than the bare value of 5, before
eventually approaching the bare value for $ h\gg U$.\par
 The fact that
the quasiparticle interaction becomes very large does not imply that the
effects of the interaction become stronger; the contrary is the case. A more
significant measure of the effect of the interactions is the product 
of $\tilde U(h)$ with the quasiparticle density of states at the Fermi level
$\rho_d(0,h)$.
The increase in $U(h)$  is more than compensated by the fall off
 of $\rho_d(0,h)$ with $h$ as the level moves away from the Fermi level. This can be seen from
the Wilson ratio $R(h)=1+\tilde\rho_d(h)\tilde U(h)$, which is also shown in Fig. 5. 
The value of $R$ decreases 
monotonically from the strong correlation value 2 to the free particle value 1
in the  extreme large field limit $h\gg U$, which implies that the product $\tilde\rho_d(h)\tilde U(h)$ is
always less than 1. The enhanced value of $\tilde U(h)$  in the strong magnetic field limit,
and the decrease as $h\to\infty$, is understandable in terms of mean field theory, where from eq. (\ref{tilu})
$\tilde U(h)$ is given by 
\begin{equation}\tilde U_{\rm mf}(h)={U\over 1-U\tilde\rho_{d,{\rm mf}}(0,h)},\label{mfu}\end{equation}
 as $\tilde\rho_{d,{\rm mf}}(0,h)$ decreases with increase of $h$ in this regime.\par
\begin{figure}[tb]
\begin{center}
\includegraphics[width=0.6\textwidth]{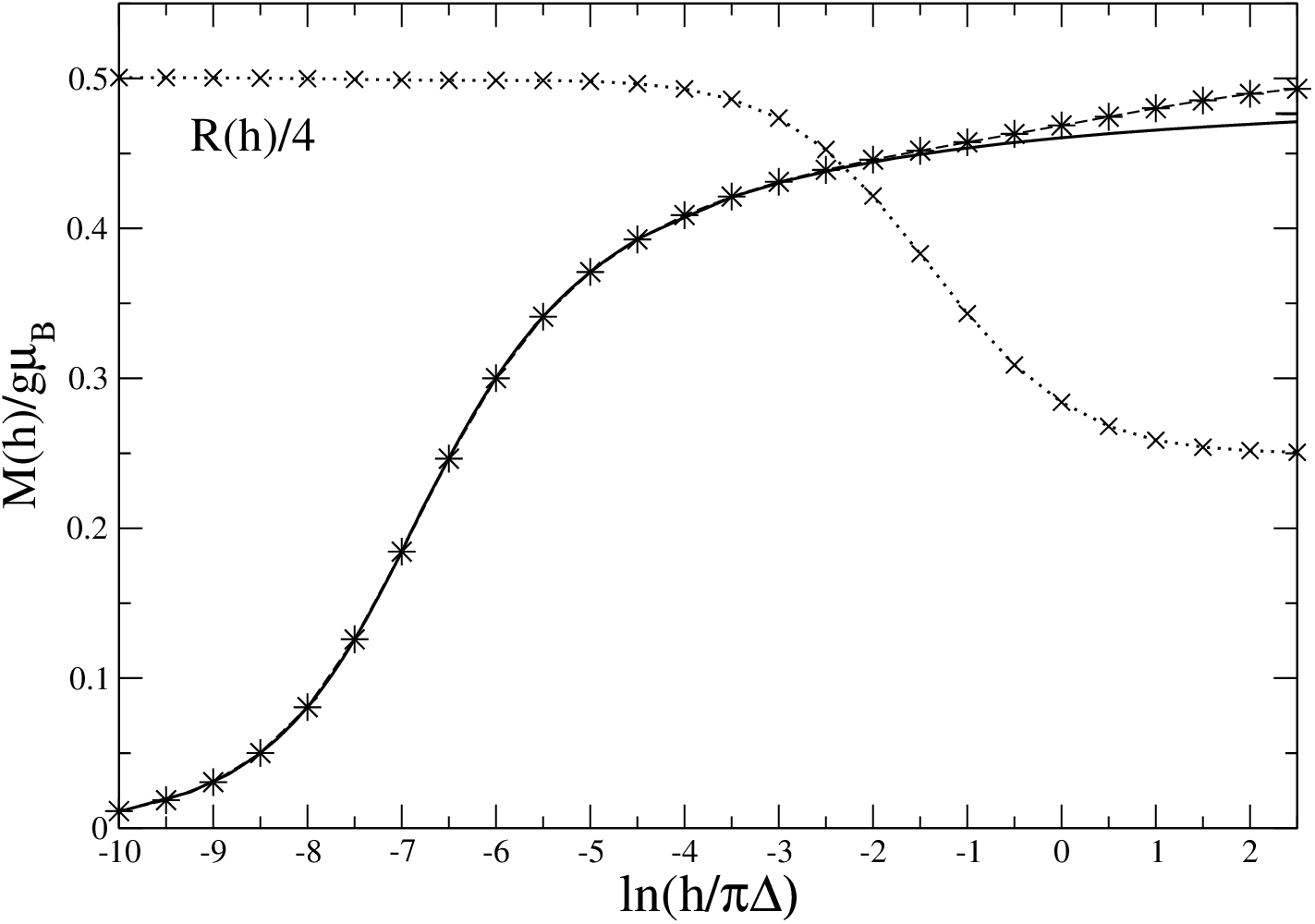}
\end{center}
\caption{The induced impurity magnetization $M(h)$ (stars and dashed curve) as a function of ${\rm
    ln}(h/\pi\Delta)$ for the symmetric Anderson  model with the same parameter set as
    used in Fig.  5 ($U/\pi\Delta=5$) compared with the same values from
    the Bethe ansatz solution\cite{afl,tw} of the s-d model (full curve). Also shown is the
    Wilson ratio $R(h)$ divided by 4 (crosses and dotted line).} 
\label{figure6}
\end{figure}
In Fig. 6 we plot the impurity magnetization  as function of the logarithm of
$h$ using the non-interacting quasiparticle expression in eq. (\ref{mag})
using the  parameters $\tilde\epsilon_d(h)$ and $\tilde\Delta(h)$ given in
Fig. 5, together with the corresponding values for an s-d model with the
same Kondo temperature\cite{tw}. 
 Also plotted in the same Fig. is $R(h)/4$. One sees that there is complete
 agreement
with the results of the s-d model until charge fluctuations begin to play a
role for
 $h\sim\pi\Delta$ (${\rm ln}(h/\pi\Delta)\sim -3$). Up to this point
  $R(h)\approx 2$ ($\tilde U(h)\tilde\rho_d(0,h)\approx 1$) independent of $h$, as is well known for the s-d model\cite{tw,rem},
but then crosses over eventually to the free electron value $R(h)=1$ 
for $h\gg\pi\Delta$. The magnetization can be seen to approach saturation 
more rapidly with $h$ than predicted from the s-d model, due to the additional
effect of the charge fluctuations.\par  
 The calculation of the magnetization from  eq. (\ref{mag}) does not
 depend on the quasiparticle interaction $\tilde U(h)$, but the formula for
 the susceptibility in eq. (\ref{rsus})  as a function of $h$ does. A check on the values of $\tilde
 U(h)$  can be made by calculating the susceptibility from eq.
 (\ref{rsus}) using $\tilde\epsilon_d(h)$, $\tilde\Delta(h)$ and  $\tilde
 U(h)$,
and comparing the result with that calculated  by numerically differentiating the results
for
$M(h)$. When this check is made complete consistency is found, confirming the
calculated values of   $\tilde U(h)$.\par

\section{Perturbation Expansion with Running Coupling Constants}
The values of
$\tilde\epsilon_d(h)$,
$\tilde\Delta(h)$ and $\tilde U(h)$ obtained in the previous section provide a
set of running coupling constants for a renormalized perturbation expansion
for the symmetric Anderson model in a magnetic field. Instead of working with
 $\tilde\Delta$ and $\tilde U$ of the model in the absence of a magnetic field,
the perturbation expansion can be carried out using parameters appropriate to
the field strength $h$. If the calculations could be carried out exactly then
it would not matter which set of coupling constants is used, as the model is
completely defined by any one set. However, for approximate calculations
in the low energy regime  the best  set in the presence of a magnetic field
 $h$ must be the set $\tilde\epsilon_d(h)$, $\tilde\Delta(h)$ and $\tilde U(h)$,
    because low order calculations in terms of these parameters
 give  asymptotically exact results as $T\to 0$.\par
We apply this approach to the calculation of the low temperature conductivity in the presence of a magnetic field for the particle-hole symmetric model $\sigma(T,h)$. The calculation is along the same lines as that used to derive the $h=0$ result in eq. (\ref{sigT})\cite{rpt1}.
All the  terms in the RPT expansion to order $\tilde U(h)^2$ are taken into account, including
the counter terms to this order. The result, when summed over the two spin components,
can be written in the form,
\begin{equation}\sigma(h,T)={\sigma_0\over {\rm cos}^2m(h)}\left\{1+
  \sigma_2(h)\left(T\over T_{\rm K}\right)^2+{\rm O}((T/T_{\rm K})^4)\right\},\label{sigTh}\end{equation}
where $m(h)=\pi M(h)/g\mu_{\rm B}$ and the coefficient $\sigma_2(h)$ is given by
$$\sigma_2(h)={\pi^3\tilde\rho_d(0,h)\tilde\Delta^2(0)\over
    {48\tilde\Delta(h)}}\left\{ 1+2\pi\tilde\Delta(h)\tilde
  U^2(h)\tilde\rho_d^3(0,h)\right\}$$
\begin{equation}={\pi^2{\rm cos}^2 m(h)\tilde\Delta^2(0)\over {48\tilde\Delta^2(h)}}\left\{1+2[R(h)-1]^2{\rm cos}^2 m(h)\right\}.\end{equation}
This result is exact to order $T^2$ if either $h$
or $U$ is set to zero, and we conjecture, that it is exact for the general case
$U\ne 0$ and $h\ne 0$. In Fig. 7 we plot the logarithm of the  coefficient
$\sigma_2(h)$ of the $T^2$ term as a function of $h$ over a range from
strongly correlated quasiparticles in weak field to bare quasiparticles in
fields $h\gg U$, 
using the renormalized parameters shown in Fig. 6 for the model with
$U/\pi\Delta=5$. In Fig. 8 we plot  $\sigma_2(h)$ for a
more realistically obtainable range of $h$, $0<h<1.5T_{\rm K}$, for the same
model where $T_{\rm K}/\pi\Delta=0.00204$.  There is a very significant decrease in this
coefficient over this magnetic field range.
\par

This RPT calculation has had to rely on the renormalized parameters derived from the
NRG energy levels. It might, however, be possible to calculate them directly
from eqs. (\ref{ren1}) and (\ref{ren2}), using an iterative RPT.
In the asymptotically large field limit, the parameters are those of the bare
model, but the interaction effects are small because the impurity is almost
completely polarized. It should be possible, therefore, to use perturbation
theory for the bare model to calculate the small renormalization effects
for large $h$, and the set of renormalized parameters for this regime.
With these new set of renormalized parameters, the RPT could be then used
to calculate the renormalized parameters for slightly smaller fields,
and enabling one to set up flow equations to continue the process to the small field regime. The perturbation effects
at each stage should be small, as the effects of only small changes in the 
magnetic field will be calculated at each stage, and there are only three
parameters, $\tilde\epsilon_d$, $\tilde \Delta$ and $\tilde U$, to consider,
as these three fully specify the  expansion.\par 
\begin{figure}[tb]
\begin{center}
\includegraphics[width=0.6\textwidth]{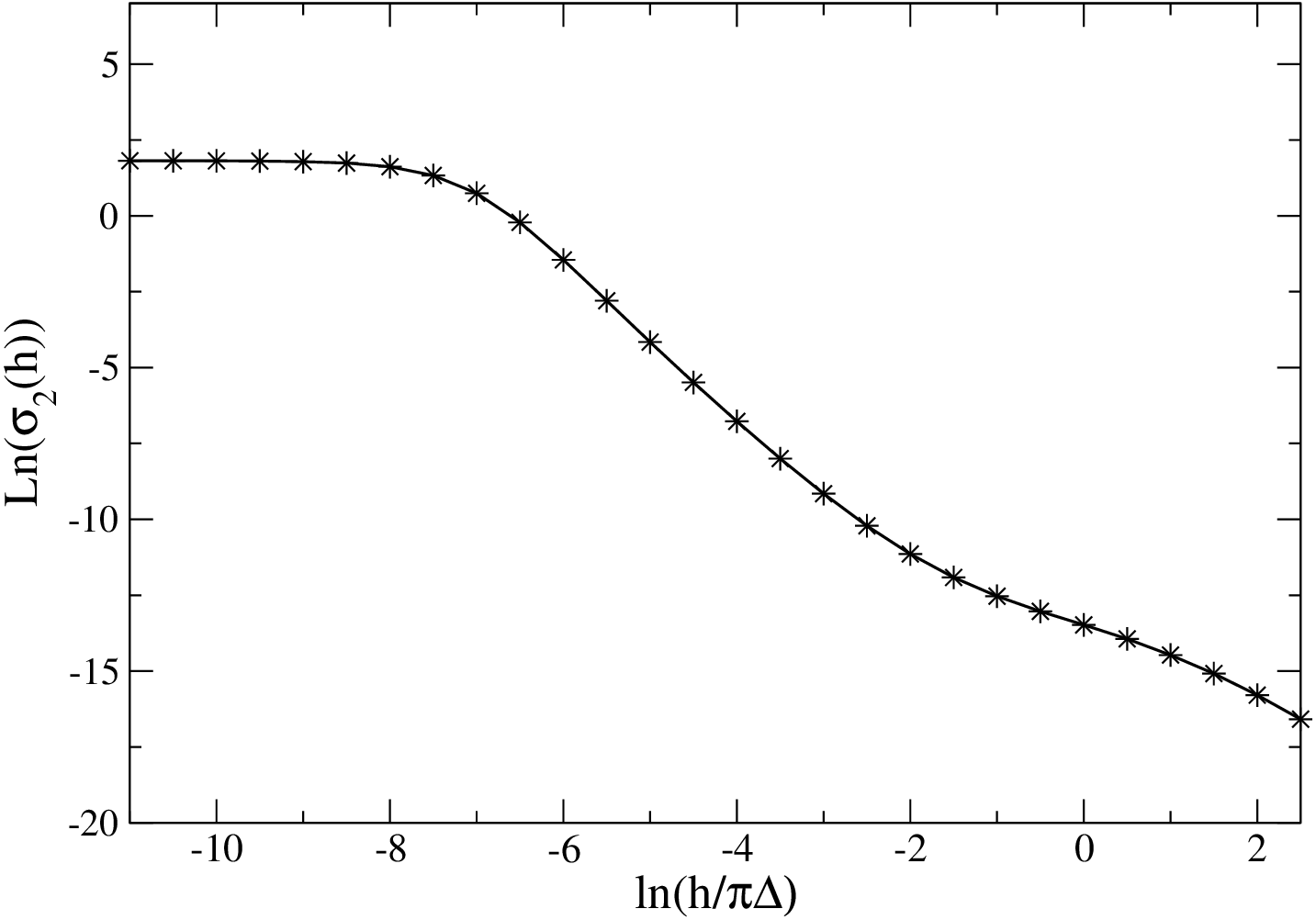}
\end{center}
\caption{ The logarithm of  $\sigma_2(h)$ (stars),  the coefficient of the
   $(T/T_{\rm
  K})^2$ term in
  the expression (\ref{sigTh}) for the low temperature conductivity
  $\sigma(T,h)$, is plotted as a function of the logarithm magnetic field
    $h$ using the renormalized parameters shown in Fig. 6 (symmetric model with $U/\pi\Delta=5$)}.
\label{figure7}
\end{figure}

\begin{figure}[tb]
\begin{center}
\includegraphics[width=0.6\textwidth]{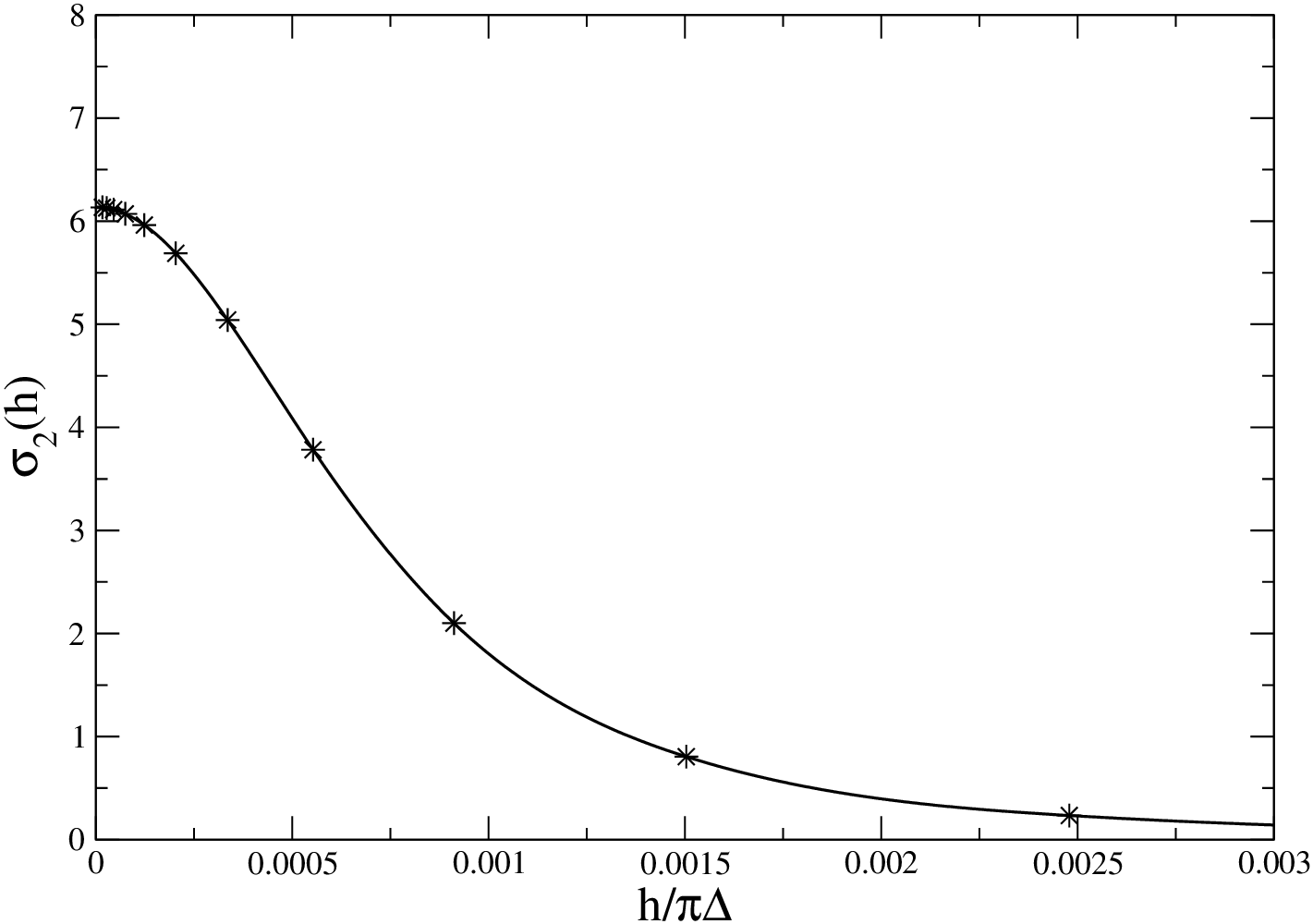}
\end{center}
\caption{  The coefficient of the $(T/T_{\rm
  K})^2$ term (stars)
  in
  the expression (\ref{sigTh}) for the low temperature conductivity
  $\sigma(T,h)$ is plotted as a function of the magnetic field
    $h$  over a range $0<h<1.5 T_{\rm K}$,
using the renormalized parameters shown in Fig. 6 (symmetric model
  with $U/\pi\Delta=5$, $T_{\rm K}/\pi\Delta=0.00204$)}.
\label{figure8}
\end{figure}

It might also be possible to do something similar as a function
of temperature starting with the high temperature limit, progressively
lowering the temperature,  hence having running parameters,  
 $\tilde\epsilon_d(T)$, $\tilde\Delta(T)$ and $\tilde U(T)$. 
We have explored this possibility by taking the parameters, 
 $\tilde\epsilon_d(N)$, $\tilde\Delta(N)$ and $\tilde U(N)$ as a function of 
$N$,  
and translating these into parameters for a temperature scale, $T_N=\eta D\Lambda^{-(N-1)/2}$, where $\eta$ is an appropriately chosen
constant of order unity\cite{hom}.
For the particle-hole symmetric case  we take the value of
$\tilde U_{pp}(N)$ (=$\tilde U_{hh}(N)$) as $\tilde U(N)$, and translate 
this, together with $\tilde\Delta(N)$ and $\tilde\epsilon_d(N)$, into
parameters appropriate for a temperature scale $T_N$, and generalize the RPT expression
for the impurity susceptibility in eq. (\ref{rsus}) to finite temperatures, 
\begin{equation}
\chi_{s}(T)={(g\mu_{\rm B})^2\over 2}\tilde\rho_{d}(0,T)(1+\tilde U(T)\tilde\rho_{d}(0,T)),\label{cht}\end{equation}
where  
$\tilde\rho_{d}(0,T)$ is the free quasiparticle contribution to the impurity
susceptibility given by
\begin{equation}
\tilde\rho_{d}(0, T)=-\int_{-\infty}^{\infty}\tilde
\rho_{d}(\omega){\partial f(\omega)\over\partial\omega} d\omega
\end{equation}
where $f(\omega)=1/(e^{\omega/T}+1)$,
and $\tilde\rho_{d}(\omega)$ is the free quasiparticle density
of states given by eq. (\ref{qpdos}).  We calculate
$\tilde\rho_{d}(0,T)$  in the Kondo
regime for $U/\pi\Delta=5.0$ at values of $T_N$, using the 
renormalized parameter $\tilde\Delta(T_N)$, with
$\tilde\epsilon_d(T_N)=0$ and   $\eta=1.2$, as is used in the NRG
evaluation of spectral densities on a scale $\omega_N$ (see for
example \cite{sak,chz}).  We then deduce $\chi_{s}(T)$ from
eq. (\ref{cht}) using $\tilde U(T_N)$. In Fig. 9 we compare the
results of this calculation with the Bethe ansatz results for the s-d
model given in reference \cite{tw}.  There is excellent
agreement with the exact Bethe ansatz results over this temperature
range. The enhanced value of $\tilde U(T)$ at very high temperatures,
as in the high field case,
is qualitatively understandable in terms of mean field theory, where
$\tilde U_{\rm mf}(T)={U/(1-U\tilde\rho_{d,{\rm mf}}(0,T)})$,
which when substituted into eq. (\ref{rsus}) gives the mean field value for
 $\chi_{s}(T)/(g\mu_{\rm B})^2=0.5\tilde\rho_{d,{\rm mf}}(0,T)/(1-U\tilde\rho_{d,{\rm mf}}(0,T))$.  
The value of $\chi_{s}(T)/(g\mu_{\rm B})^2$ in the extreme high
temperature range corresponds to that of the free bare model, $1/8T$.
It can be seen from Figs. 1 and 2 that in the smaller  $N$ range there is no
unique prediction for $\tilde U(N)$, so this calculation is not well defined. 
Nevertheless, the agreement with the Bethe ansatz results is remarkable and
it does indicate that an RPT expansion with temperature dependent parameters
might be possible. Such a perturbation expansion, being a general technique, would
have a potentially wide application to other  strongly correlated systems.   
\begin{figure}[tb]
\begin{center}
\includegraphics[width=0.6\textwidth]{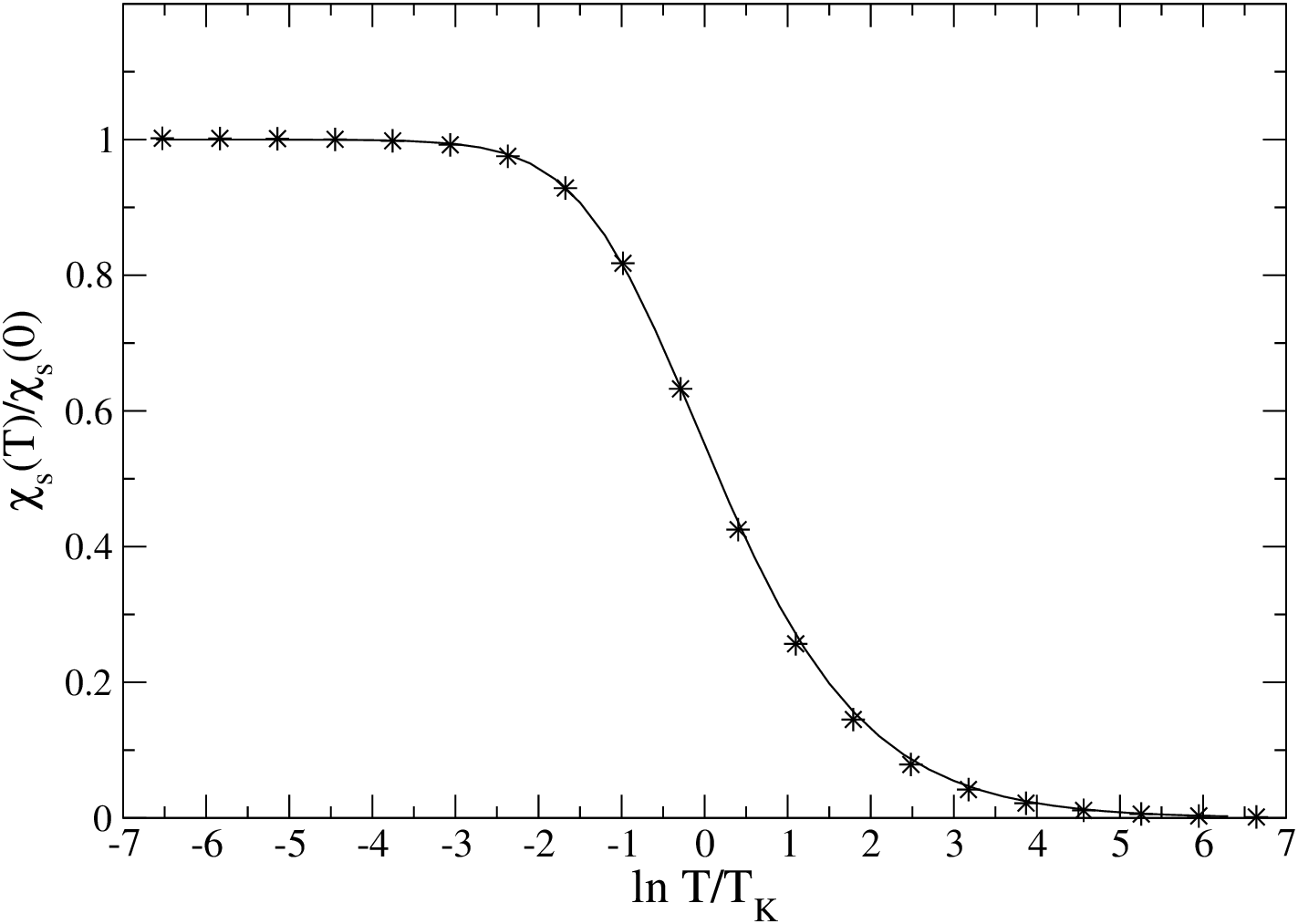}
\end{center}
\caption{The ratio of $\chi_s(T)/\chi_s(0)$ (stars) deduced from eq. (\ref{cht})
using temperature dependent parameters $\tilde\Delta(T)$ and $\tilde U(T)$, for
the symmetric model with $U/\pi\Delta=5$ as a function of the temperature $T$.
The full curve is that derived from the Bethe ansatz solution of the s-d model\cite{tw}.}
\label{figure9}
\end{figure}

\section{Conclusions}

We have shown that low energy or strong coupling NRG fixed point of the s-d and Anderson models
can be analysed as a renormalized version of the Anderson model.
This analysis has the advantage that the Fermi liquid aspects, and the 1-1  
correspondence of the single particle excitations with those of the non-interacting system,
is brought out  clearly. The three renormalized parameters, $\tilde\epsilon_d$,
$\tilde\Delta$ and $\tilde U$, can be used to specify  completely a renormalized perturbation expansion, which is applicable on all energy scales. \par
With a magnetic field present, renormalized parameters,   $\tilde\epsilon_d(h)$,
$\tilde\Delta(h)$ and $\tilde U(h)$ have been calculated as a function of the magnetic field strength $h$. Using these parameters in the renormalized perturbation theory, 
we have derived an expression for the low temperature conductivity as a function of magnetic field strength.\par

\section*{Acknowledgement}
  I wish to thank the EPSRC (Grant GR/S18571/01) for
  financial support, and A. Oguri, D. Meyer and W. Koller for helpful
discussions and cooperation on many aspects of the work described here.


\begin{thebibliography}{99}
\bibitem{com} The full definition of this ratio is $R=4\pi^2\chi_s/3(g\mu_{\rm B})^2\gamma$.\par
\bibitem{rem} The Fermi liquid relation $4\chi_s/(g\mu_{\rm B})^2+\chi_c=6\gamma/\pi^2$
can be seen to hold from eqs. (\ref{rgam}) and (\ref{rsus}). In the localized regime
$\chi_c$ is negligible and it then follows from \cite{com} that $R=2$.\par
\end{thebibliography}
\end{document}